\documentclass{jfm}
\usepackage{graphicx}
\usepackage{epstopdf, epsfig}
\usepackage[normalem]{ulem}
\usepackage[shortlabels]{enumitem}
\usepackage{CJK}
\usepackage{amssymb,euscript,latexsym,amsmath}
\usepackage{graphicx}
\usepackage{color}
\usepackage{epstopdf}
\usepackage{setspace}
\usepackage{subfigure}
\usepackage{chngpage}
\usepackage{textgreek}
\usepackage{tikz}
\usepackage{url}

\def\f{{\rm f}}
\def\b{{\rm b}}
\def\w{{\rm w}}
\def\ii{{\rm i}}
\def\Real{{\rm {Re}}}
\def\Imag{{\rm {Im}}}

\shorttitle{Antiphase tail flexion enhances swimming speed and efficiency
}
\shortauthor{H.Hang, S. Heydari, J. H. Costello and E. Kanso}

\title{Active tail flexion in concert with passive hydrodynamic forces improves swimming speed and efficiency}

\author{Haotian Hang\aff{1}, Sina Heydari\aff{1}, John H. Costello\aff{2,3}, \and Eva Kanso\aff{1} \corresp{\email{Kanso@usc.edu}}
 }

\affiliation{\aff{1}Department of Aerospace and Mechanical Engineering,  University of Southern California, 854 Downey way, Los Angeles, California 90089, USA
\aff{2}Biology, Providence College, Providence, Rhode Island 02918, USA
\aff{3}Whitman Center, Marine Biological Laboratory, Woods Hole, Massachusetts 02543, USA}

\begin{document}

\maketitle

\begin{abstract}
Fish typically swim by periodic bending of their bodies. Bending seems to follow a universal rule; it occurs at about one-third from the posterior end of the fish body with a maximum bending angle of about $30^o$. However, the hydrodynamic mechanisms that shaped this convergent design and its potential benefit to fish in terms of swimming speed and efficiency are not well understood. It is also unclear to what extent this bending is active or follows passively from the interaction of a flexible posterior with the fluid environment. Here, we use a self-propelled two-link model, with fluid-structure interactions described in the context of the vortex sheet method, to analyze the effects of both active and passive body bending on the swimming performance. We find that passive bending is more efficient but could reduce swimming speed compared to rigid flapping, but the addition of active bending could enhance both speed and efficiency. Importantly, we find that the phase difference between the posterior and anterior sections of the body is an important kinematic factor that influences performance, and that active antiphase flexion, consistent with the passive flexion phase, can simultaneously enhance speed and efficiency in a region of the design space that overlaps with biological observations. Our results are consistent with the hypothesis that fish that actively bend their bodies in a fashion that exploits passive hydrodynamics can at once improve speed and efficiency. 
\end{abstract}

\begin{keywords} 
Swimming, hydrodynamics, flexibility, bending rules, body deformations, vortex-sheet model
\end{keywords}

\section{Introduction}\label{sec:intro}
%----
\begin{figure*}
\centering
\includegraphics[scale=1]{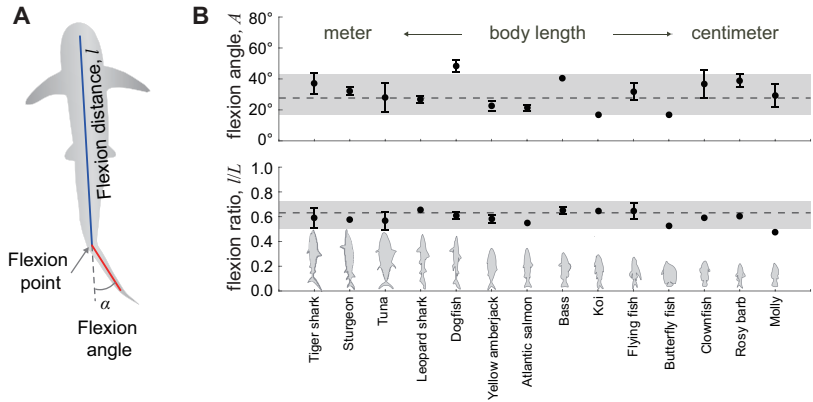}
\caption[]{\footnotesize \textbf{Bending rules of swimming fish.} (A) Schematic of flexion parameters in fish with flexion distance $l$, ratio $l/L$ and maximum angle defined as in \cite{Lucas2014}. (B) Flexion ratio $l/L$ and maximum flexion angle $A$  observed in different fish species; data taken from \cite{Lucas2014}. The observed flexion ratio and angles are fairly consistent among different fish species, despite large variations in length scales. 
% (C) Comparison of fish flexion parameters (black dots) and regions of optimal performance predicted by our model: blue region corresponds to 2$\times$ enhancement in swimming speed, and pink region corresponds to $3\times$ enhancement in swimming efficiency, both compared to a rigid swimmer of the same total length with no flexion. The contour line enclosed a region of $6\times$ enhanced efficiency.
}
\label{fig:biodata}
\end{figure*}
%----
Millions of years of natural selection endowed fish with remarkable abilities to swim efficiently compared to underwater man-made propulsors~\citep{Sfakiotakis1999,Lauder2005}. 
Several fish use body and caudal fin (BCF) deformations for propulsion. Details of BCF deformations have been used to classify fish swimming modes \citep{Sfakiotakis1999,Shadwick2005,Low2010,Smits2019}. 
Invariably, in most BCF swimming modes, anterior to posterior bending of the fish body seems to play an important role in swimming efficiency, and it is often linked to increased flexibility towards the fish tail or caudal peduncle \citep{Combes2003,Tytell2004,Tytell2010,Tytell2016,Gemmell2016}. 

In an effort to document the bending rules of fluid-based propulsors, both aerial and aquatic, \cite{Lucas2014} collected morphometric data of the flexion parameters across length scales and animal taxa. They identified two parameters to characterize the bending behavior: flexion ratio $l/L$ between the flexion distance $l$ and total length $L$ of the propulsor and maximum flexion angle $A$. They found that flexion ratio and maximum flexion angle of all surveyed animals, including fish, clustered in a limited design space: bending occurs at about $70 \%$ of body length at maximum flexion angle of about $30 ^\circ$, as shown in Figure~\ref{fig:biodata} for swimming fish based on the data collected in \cite{Lucas2014}. 
It is unclear the extent to which this anterior-to-posterior bending is active or whether it follows passively due to the interaction of a flexible posterior with the fluid motion. 
Either way, these findings raise the question of whether hydrodynamics could have provided a selective force for driving this convergent bending design.

To address these questions, we analyze the influence of bending on the swimming speed and efficiency of a simplified fish model that consists of anterior and posterior sections connected via a rotational joint at the flexion point (see Figure~\ref{fig:biodata}A). The fish anterior undergoes  periodic planar pitching while the posterior either (i) moves in synchrony with the anterior as if the two parts were a single rigid body, (ii) bends actively at distinct amplitude and phase relative to the fish anterior, or (iii) bends passively due to interactions with the flow generated by the fish anterior. We find that swimming with passive bending could be more efficient than rigid flapping but at the cost of diminished swimming speed. Active bending provides more possibilities to alter the swimming performance through not only the flexion ratio and maximum flexion angle as reported in~\cite{Lucas2014} but also the phase difference $\phi$ between the flapping motions of the anterior and posterior parts. Importantly, we find that antiphase anterior-to-posterior flexion can simultaneously enhance swimming speed and efficiency in a region of the design parameter space $(l/L, A)$. Flexion ratios and angles that lead to significant improvements in speed and efficiency overlap with the observations of real fish reported in \cite{Lucas2014}. We analyze in depth the hydrodynamic mechanisms underlying these improvements in swimming performance. 

Details of the flow field around swimming fish have received a great deal of attention. 
Several studies used particle image velocimetry to measure the flow field around live fish and analyze the interplay between body deformations and thrust production; see, e.g.,~\cite{Muller1997, Muller2002, Liao2003, Tytell2004, Gemmell2016}. 
In-silico models of various degrees of fidelity to fish morphology and kinematics have also been used to examine the offsets of body deformations on swimming speed and efficiency~\citep{Eldredge2006, Kern2006, Tytell2010, Eloy2013}. 
Importantly, several experimental and numerical studies have shown that plates and foils undergoing pitching or heaving motions provide good approximations of the fluid-structure interactions in swimming fish \citep{Blondeaux2005,Wen2013,Lauder2011,Menon2019}; including the reverse von K{\'a}rm{\'a}n vortex wake left behind swimming fish and flapping foils \citep{Taneda1965,Triantafyllou1993}. 

A variety of fluid-structure interaction models have been proposed to analyze the effect of body flexibility on bending in flows; see, e.g.,~\cite{Shelley2011} and references therein. Here, we present a brief  literature review  focused on this topic.
\cite{Heathcote2007} conducted experiments on a flapper with a rigid leading edge and flexible tail fixed in a water channel and found that flexibility can enhance both efficiency and thrust production. \cite{Eldredge2008} and \cite{Eldredge2010} numerically simulated the flapping motion of articulated rigid links and found that joint flexibility can reduce the power required for flapping.  
\cite{Alben2008} used a filament of uniform flexibility to model the tail of swimming fish in the context of the vortex sheet method and predicted enhancement in efficiency rather than thrust when choosing parameters (dimensionless rigidity and reduced pitching frequency) that are consistent with biological data. 
\cite{Quinn2015} conducted a large set of experiments on two-dimensional pitching and heaving flexible plates at various stiffness values, kinematic parameters, and incoming flow speeds. By combining grid search and gradient-based optimization methods, they found that optimizing the pitching angle with heaving can almost double the propulsor efficiency compared to heave-only motions. 
\cite{Hoover2018} conducted simulations combining three-dimensional Navier-Stokes equation with one-dimensional Euler–Bernoulli beam theory to analyze the motion of heaving flexible plates, and identified local peaks in swimming speed over a parameter space consisting of the beam material property and heaving frequency.
Also using three-dimensional simulations of pitching plates of uniform flexibility, \cite{Dai2012} found that the phase delay $\phi$ between the leading and trailing edge of the plate decreases with increasing stiffness $\kappa$. For large stiffness, the plate moves in no-neck mode (in-phase in our notation), in which thrust production is close to that of a rigid pitching plate with similar trailing edge displacement. 

The effect of uniform flexibility on swimming speed, thrust generation, swimming energetics, and stability have been analyzed in numerous other experimental and computational studies; see, e.g., \cite{Shoele2016,Feilich2015,Michelin2009,Miao2006,Combes2001,Hua2013,Wang2020,Ryu2019}. 
Specifically,~\cite{Liu1997,Heathcote2008,Tangorra2010} have indicated that flexibility could lessen or prevent thrust production.
Flexible propulsors have also been widely used in man-made biomimetic underwater autonomous vehicles (UAV); see, e.g., \cite{Fujiwara2017,Katzschmann2018,Gibouin2018,Zhu2019,White2021}. Most notable is the Tunabot design of  
\cite{Zhu2019} and \cite{White2021} which mimics the shape and bending kinematics of yellowfin tuna (\textit{Thunnus albacares}) and Atlantic mackerel (\textit{Scomber scombrus}). 

While most studies have focused on uniformly flexible bodies, \cite{Combes2003} and \cite{Lucas2014} noted that the stiffness along the fish body is not uniform, but decreases towards the tail, and that the propulsor becomes highly flexible at the flexion point of the body; see Figure~\ref{fig:biodata}(A). 
To explore the effects of non-uniform flexibility on efficiency and thrust production, \cite{Lucas2015} considered flexible plates of inhomogeneous stiffness undergoing heaving and pitching motions in a water tunnel and found that non-uniform stiffness can improve thrust production, and that in order to achieve optimal propulsion, the morphologic factor (flexion ratio) and kinematic factor (motion type and motion parameters) should be considered simultaneously. 
\cite{Vincent2020a,Vincent2020b} analyzed the effect of non-uniform flexibility on flight performance in the context of a tumbling wing model, and found that wing tip flexibility that follows the empirical rules reported in \cite{Lucas2014} leads to improved flight performance.

In this paper, we use a simplified two-link fish model to analyze the influence of active and passive bending on the swimming speed and efficiency. Two link models are commonly used to study the effect of flexibility on the performance of flapping bodies \citep{Eldredge2010, Wan2012,Li2015}. We solve for fluid-structure interactions in the context of the vortex sheet method as described in Section~\ref{sec:formulation}. 
The vortex sheet model has been used extensively to analyze problems of fluid-structure interactions, including ring formation at the edge of a circular tube~\citep{Nitsche1994} and wakes of oscillating plates~\citep{Jones2003,Sheng2012}, falling cards~\citep{Jones2005}, flapping flexible flags~\citep{Alben2008,Alben2009}, swimming plates~\citep{Wu1971}, hovering flyers~\citep{Huang2016, Huang2018} and schooling of swimming plates~\citep{Heydari2020}. 
Here, we use the implementation of \cite{Nitsche1994}. 
In Section~\ref{sec:result1}, we report the effects of both active and passive bending on the swimming performance of sinusoidally pitching swimmers compared to rigid flapping. 
In Section~\ref{sec:con}, we discuss these findings in the context of existing work and highlight the implications of our results on the design of underwater autonomous vehicles.

\section{Problem formulation}\label{sec:formulation}
%----
\begin{figure*}
\centering
\includegraphics[scale=1]{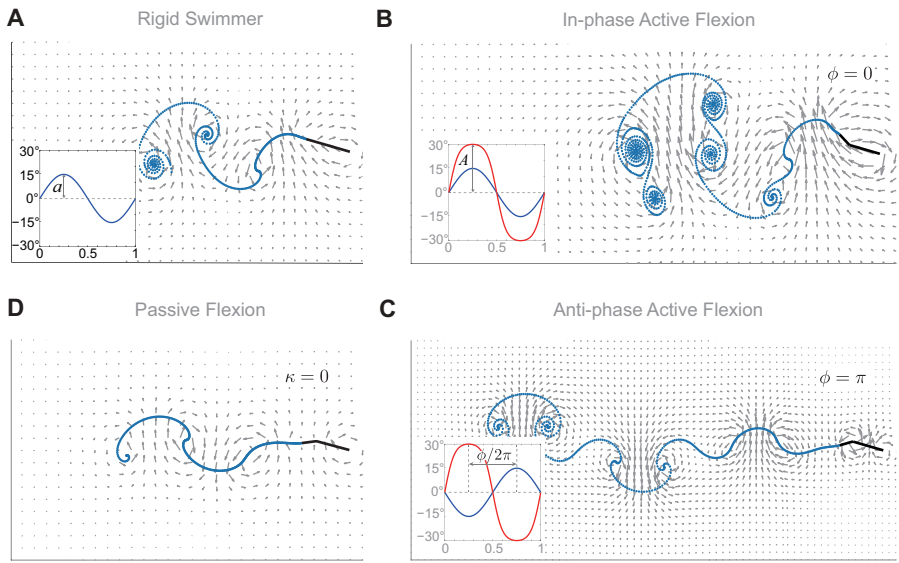}
\caption[]{\footnotesize \textbf{Wake and flow velocity of free swimmers.} (A) Rigid swimmer undergoing periodic pitching (inset) of period $T=1$ and amplitude $a=15^\circ$.  (B,C) Active bending with both anterior and posterior sections undergoing periodic pitching albeit at different amplitude and phase. The anterior follows the same pitching motion (blue line in inset) as the rigid swimmer while the relative rotation of the posterior follows a prescribed Jacobi elliptic sine function (red line in inset) with flexion amplitude $ A = 30 ^ \circ$, flexion ratio $l/L=0.7$, elliptic modulus $M=0.9$, and phase  $\phi= 0$ (in-phase)  and $\phi=\pi$ (antiphase) in B and C, respectively. (D) Passive bending of posterior section while anterior section follows the same prescribed pitching as the rigid swimmer. Joint parameters are set to $\kappa =0$ and $c=1$.
The dissipation time is set to be $T_{\rm{diss}}=1.65T$ in (A-C), $\sqrt{2.09}T$ in (D).}
\label{fig:flowfield}
\end{figure*}
%----

We model the flexible swimmer as a planar two-link body of total length $L$, negligible thickness $e\ll L$, and total mass per unit depth $m=\rho eL$, where $\rho$ is both the swimmer and fluid density, assuming a neutrally-buoyant fish. 
The flexion point indicates where the anterior link (of length $l$) is joined to the posterior link; see Figure~\ref{fig:biodata}(A). 
The anterior link undergoes sinusoidal pitching motion $\theta_a(t) = a\sin(2\pi f t)$, where $\theta_a$ is the angle relative to the swimming direction, taken to be parallel to the $x$-axis. Here, $a$ is the flapping amplitude, $f={2 \pi}/T$ the flapping frequency, and $T$ the flapping period.
When the posterior is connected rigidly to the anterior link at zero flexion, the posterior motion $\theta_p(t)$ is equal to $\theta_a(t)$ and the flexion angle $\alpha = \theta_p - \theta_a$ is identically zero for all time. The two links form a single rigid plate (Figure~\ref{fig:flowfield}A) whose swimming motion due to sinusoidal pitching has been extensively analyzed~\citep{Jones2003,Sheng2012,Moored2019,Heydari2020,Labasse2020}. To explore the effects of body bending on swimming, we consider two cases: (i)  
active bending where the flexion angle $\alpha(t)$ is controlled by the swimmer, and (ii) passive bending where the flexion angle $\alpha(t)$ is dictated by the physics of fluid-structure interactions.

When the two-link swimmer bends actively, we allow the anterior link to have a phase advantage of magnitude $\phi$ relative to the flapping motion of the posterior link. At $\phi = 0$, both anterior and posterior links flap in phase and the swimmer bends in the direction of flapping (Figure~\ref{fig:flowfield}B); for $\phi = \pi$, they flap antiphase resulting in bending in the opposite direction to the anterior pitching motion (Figure~\ref{fig:flowfield}C). The  flexion angle $\alpha(t)$ follows a Jacobi elliptic sine function $\alpha(t)=A \operatorname{sn}(4 K f t, M)$, where $A$ is the maximum flexion angle and $M$ is the elliptic modulus that controls the shape of the elliptic sine function. As $M \to 0$, the elliptic sine function tends to a sinusoidal function and as $M \to 1$, it approaches a square wave shape. The parameter $K$ is introduced to ensure that the flapping frequency of the posterior link is the same as that of the anterior link ($K$ is related to the elliptic modulus $M$ via the elliptic integral and it is given by $K = \mathrm{ellipticF}(\pi / 2, M)$ in Matlab). In this paper, without other specification, we fix $M=0.9$ and explore the effects of the anterior-to-posterior phase difference $\phi$, flexion ratio $l/L$, and maximum flexion angle $A$ on the swimming performance. 
% The effect of elliptical modulus $M$ is discussed in Appendix~\ref{app:M}.

We write the equations governing the self-propelled motion of the two-link swimmer in non-dimensional form. To this end, we  scale all parameter values using $L/2$ as the characteristic length scale, $1/f$ as the characteristic time scale, and $\rho (L/2)^2$ as the characteristic mass per unit depth. Accordingly, velocities are scaled by $Lf/2$, forces by $\rho f^2 (L/2)^3$, moments by $\rho f^2 (L/2)^4$, and power by $\rho f^3 (L/2)^4$. 
The equation of motion governing the free swimming $x(t)$ is given by Newton's second law
%--
\begin{equation}
m \ddot{x}=  F_x - D_x .
\label{eq:eom}
\end{equation}
%--
Here, $F_x$ and $D_x$ denote the components of the hydrodynamic  pressure and drag forces $F$ and $D$ in the swimming direction. Specifically, $F$ is the total force  normal to the swimmer due to hydrodynamic pressure and $D$ is the force tangential to the swimmer due to skin drag. The hydrodynamic pressure force is calculated in the context of the inviscid vortex sheet model~\citep{Nitsche1994, Huang2016, Huang2018, Heydari2020}, and the drag force $D$ is introduced to emulate the effect of fluid viscosity~\citep{Fang2016,Ramananarivo2016}. A brief overview of the vortex sheet method and its numerical implementation is given in Appendix~\ref{app:VSM} and \ref{app:numerics}. Detailed expressions of the fluid forces acting on the swimmer are given in Appendix~\ref{app:forces}. 

When the swimmer bends passively, the relative rotation $\alpha(t)$ of the posterior end is not prescribed a priori and follows from the physics of fluid-structure interactions. Considering that the rotational joint at the flexion point is equipped with a torsional spring of stiffness $\kappa$ and damping coefficient $c$, we write the equation governing the rotational motion of the posterior link
%--
\begin{equation}
I_{p}(\ddot{\theta}_a + \ddot \alpha) +  c \dot{\alpha} + \kappa \alpha =  M_{p} + M_{\rm inertia}, 
\label{eq:eomalpha}
\end{equation}
%--
where $I_{p}$ and $M_p$ are the moment of inertia and hydrodynamic moment acting on the posterior link about the flexion point, $M_{\rm{inertia}}$ is an inertial moment that arises because the flexion point about which the moments are balanced is moving;  see details in Appendix~\ref{app:forces}. 

To assess the swimming performance of the two-link swimmer, we introduce five metrics: the period-average swimming speed $U=\int_t^{t+T}{\dot x dt}$ at steady state, the thrust force $F_x$, the period-average input power $P=\int_t^{t+T}{ P(t) dt}$ required to maintain the prescribed flapping motions, and the propulsion efficiency $mU^2/2PT$ defined as the kinetic energy of the swimmer divided by the input work over one flapping period; see Appendix~\ref{app:forces} for more details. 

%----
\begin{figure*}
\centering
\includegraphics[scale=1]{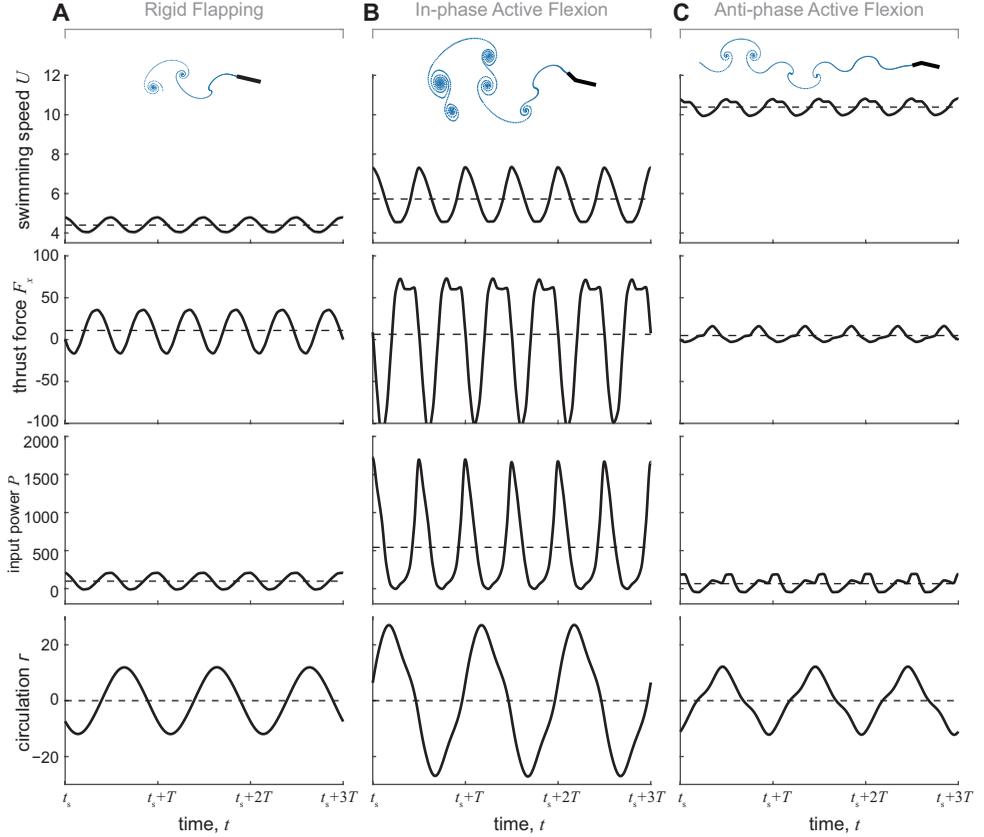}
\caption[]{\footnotesize \textbf{Active anterior-to-posterior bending of free swimmers.} Time-dependent speed, thrust, input power and circulation for (A) rigid swimmer undergoing pitching at $a=15 ^ \circ$, $T=1$, active flexion (B) at phase difference $ \phi=0$, (C) at phase difference $ \phi=\pi$. In (B) and (C), flexion ratio $l/L=0.7$ and flexion angle $A=30 ^ \circ$. 
Solid lines represent the instantaneous values and dashed lines represent time-period averages. The results are shown after the swimmers have reached steady state $t_{\rm{s}}=15T$. The dissipation time is set to be $T_{\rm{diss}}=1.65T$. 
}
\label{fig:compare}
\end{figure*}
%----
%----
\begin{figure*}
\centering
\includegraphics[scale=1]{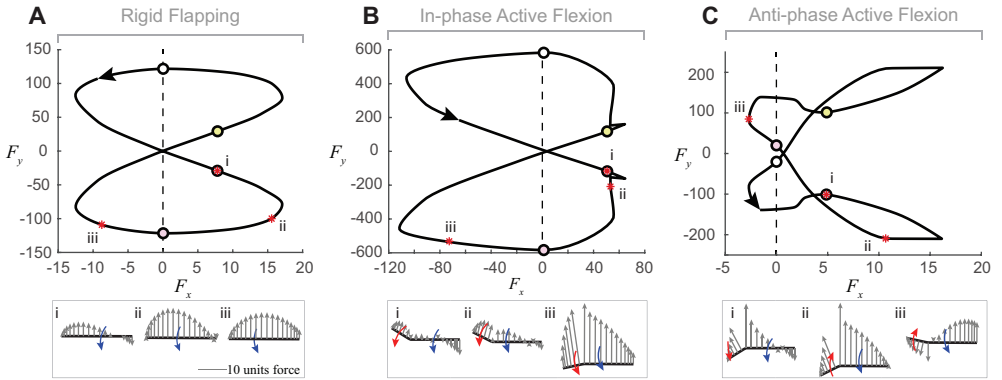}
\caption[]{\footnotesize \textbf{Active anterior-to-posterior bending can minimize lateral  forces and negative thrust.} Force hodograph of (A) rigid flapping, (B) in-phase flexion, (C) antiphase flexion for the same cases shown in Figure~\ref{fig:compare}. The arrow indicates the direction of time. The white, blue, red, yellow points are: $t/T= 0$, $0.25$, $0.5$, $0.75$. Bottom row shows force distribution at the instants indicated in red stars. 
Blue arrows represent pitching direction of the anterior link motion while red arrows  of the posterior link.
}
\label{fig:hodograph}
\end{figure*}
%----

\section{Results} \label{sec:result1}
We compare the free swimming that results from flapping while undergoing active and  passive bending to that of rigidly flapping. All swimmers have the same total length $L$ and undergo the same sinusoidal pitching motion about their leading edge $\theta_a = a\sin(2\pi f t)$ with $a= 15^o$ and $f=1$, albeit exhibiting distinct bending patterns. Figure~\ref{fig:flowfield} shows snapshots of the wake represented by the free vortex sheet and velocity field generated by a swimmer undergoing (A) rigid flapping, (B) in-phase active bending with flexion amplitude $A = 30^o$ and flexion ratio $l/L = 0.7$, (C) antiphase active bending at the same flexion amplitude and ratio, and (D) passive bending. All snapshots are taken at the same instant in the flapping cycle (at $0.25T$ after steady state has been reached). 
Compared to the rigid swimmer, in-phase flexion produces wider wakes and larger leading edge circulation and instantaneous flow speeds, while antiphase flexion is characterized by a leaner, longer wake with weaker leading edge circulation and lower flow speeds. 
The main features of the instantaneous flow during antiphase flexion, namely, the leaner wake and weaker leading edge circulation are also observed in passive bending of the swimmer body. These flow features are important for the hydrodynamic forces exerted on the swimmer as discussed later.

%----
\begin{figure*}
\centering
\includegraphics[scale=1]{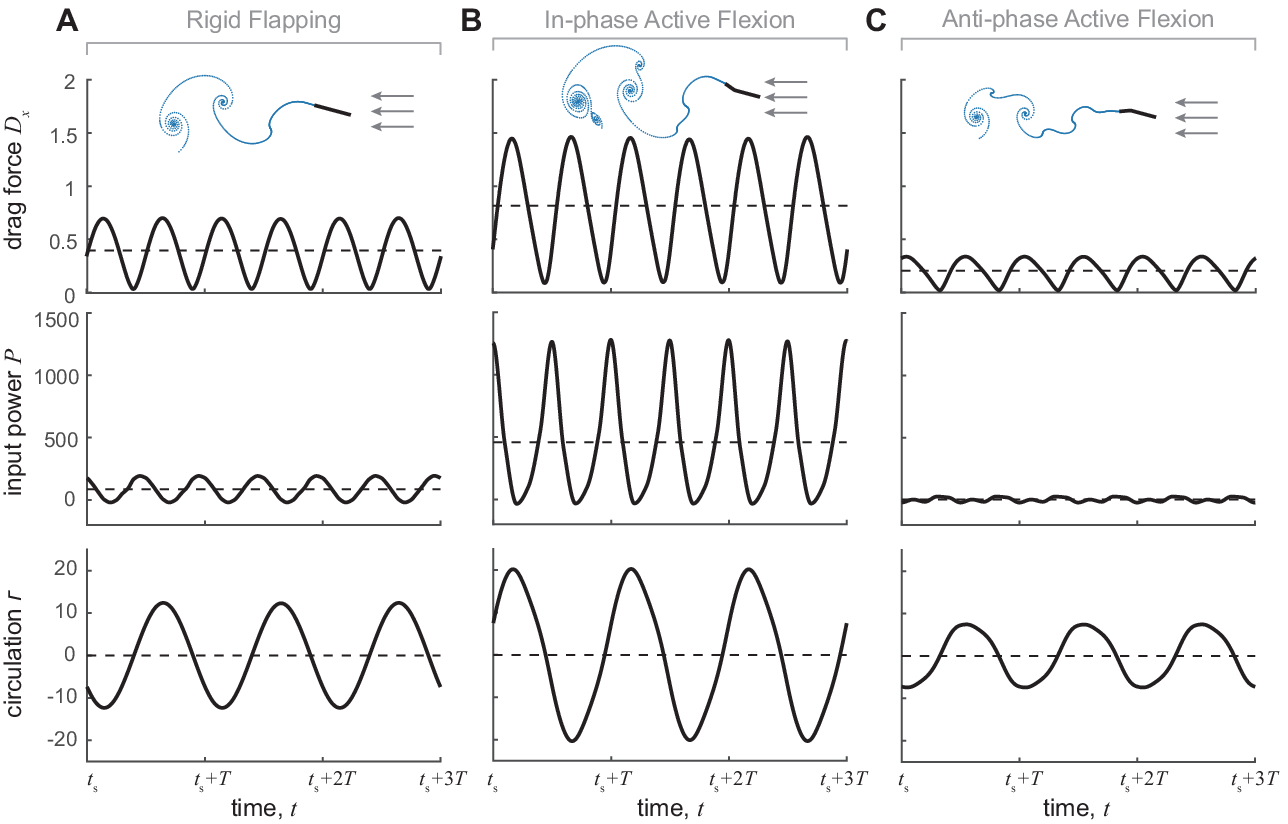}
\caption[]{\footnotesize \textbf{Active anterior-to-posterior bending of swimmers fixed in oncoming flow.} Time-dependent drag force, input power, and wake circulation for (A) rigid swimmer,  (B) in-phase active flexion ($\phi=0$), (C) antiphase active flexion ($\phi=\pi$). In all cases, the swimmer is fixed in a uniform oncoming flow at $U=9$. 
Results are shown after the swimmers have reached steady state $t_{\rm{s}}=11T$. From top to bottom, drag force, input power and circulation of wake are shown. Solid lines represent the instantaneous values and dashed lines represent time-period averages. The dissipation time is set to be $T_{\rm diss}=1.65T$.  
}
\label{fig:compare2}
\end{figure*}
%----

We quantitatively evaluate the steady state motion of the rigid and actively-bending swimmers in Figure~\ref{fig:flowfield}(A-C). In Figure~\ref{fig:compare}, from top to bottom, we report the instantaneous (solid lines) and period-average (dashed lines) values of the swimming speed, thrust force, input power, and circulation. 
On average, rigid flapping produces the lowest swimming speed while antiphase flexion the highest. Fluctuations around the average swimming speed are smallest for the swimmer undergoing antiphase flexion. 
The discrepancy in average swimming speeds between the three flapping modes is surprising at first sight given that the average values of the thrust force are comparable.  However, a closer look at the instantaneous thrust shows that the swimmer undergoing antiphase flexion hardly experiences negative thrust over its flapping cycle. In-phase flexion leads to negative thrust of high magnitudes over larger subintervals of the flapping cycle, as highlighted further in Figure~\ref{fig:hodograph}.
Consequently, the required input power for in-phase flexion is largest compared to both rigid flapping and antiphase flexion.
This is also true of the overall wake circulation.  It is worth noting that, by Kelvin's circulation theorem, circulation around the leading edge must be equal to the overall circulation in the wake. Therefore, compared to rigid flapping, in-phase flexion increases the circulation around the leading edge of the swimmer while antiphase flexion decreases leading edge circulation as noted qualitatively in Figure~\ref{fig:flowfield}. 

%----
\begin{figure*}
\centering
\includegraphics[scale=1]{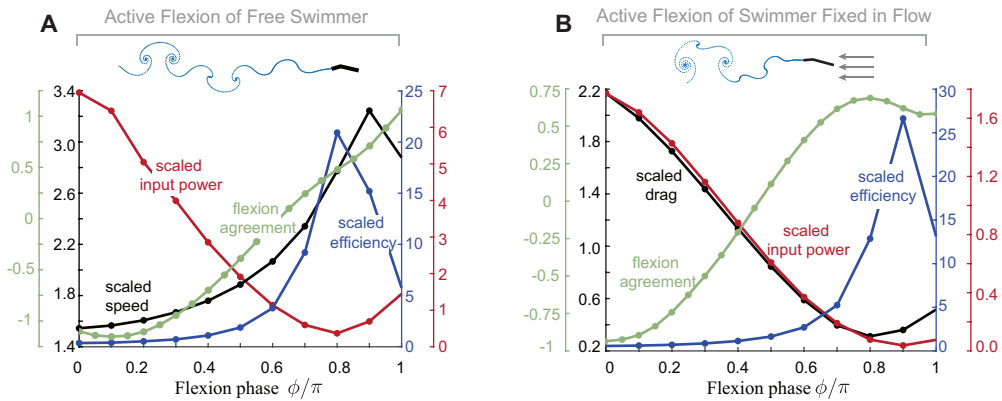}
\caption[]{\footnotesize \textbf{Performance of active anterior-to-posterior bending as a function of phase.} 
(A) Free swimmer: swimming speed (black),  power requirement (red), and efficiency (blue) are scaled by the corresponding values of a rigid swimmer.  (B) Swimmer fixed in oncoming flow of uniform speed $U$: drag force (black), power requirement (red) and efficiency (blue) are scaled the corresponding values of a pitching rigid plate fixed oncoming uniform flow. Parameters values are set to $a=15^\circ$, $l/L=0.7$, $A=20^\circ$, $M=0.9$, $U=9$. Flexion agreement parameter $Z$ between the relative velocity of an actively bending posterior and the fluid velocity generated by a passively bending swimmer at zero stiffness $\kappa = 0$ (green) as a function of phase $\phi$ during (A) free swimming  and (B) holding station in oncoming flow $U=5$. 
}
\label{fig:mix}
\end{figure*}
%----

The results in Figure~\ref{fig:compare} indicate that the swimmer undergoing antiphase flexion achieves higher swimming speed at lower power requirement and energetic cost. To elucidate the hydrodynamic forces at play, we report in Figure~\ref{fig:hodograph} the force hodograph defined as a plot of the lateral pressure force $F_y$ versus thrust $F_x$ acting on each swimmer.   
Snapshots of the distribution of hydrodynamic pressure forces along the swimmers are depicted in bottom row of Figure~\ref{fig:hodograph}, and indicate that the $x$-component of the forces on the anterior and posterior sections during antipase flexion act opposite to each other as in a tug-of-war, leading to overall reduction in thrust values. 
Importantly, antiphase flexion also reduces the lateral force and negative thrust, with negative thrust experienced only over a  small subinterval of the flapping period, as noted earlier.
In contrast, in-phase flexion significantly increases the lateral force and negative thrust.

To explain the effect of flexion on the lateral force experienced by the swimmer, it is instructive to re-examine the flow field around the rigid and actively-bending swimmers in Figure~\ref{fig:flowfield}(A-C). 
A large leading edge vortex (LEV) is known to generate large lift in flapping flight; see, e.g.,~\cite{Ellington1984,Dickinson1999}. In swimming, larger leading edge circulation creates larger lateral force, which explains why, compared to rigid flapping, in-phase flexion increases the lateral force acting on the swimmer while antiphase flexion decreases it. 
Lift is beneficial for flight but large lateral forces are detrimental to swimming speed, as noted in~\cite{Drucker2000} for fish and recapitulated here in the context of our swimmer model.

To further analyze the difference in the swimming performance between rigid flapping and flapping with inphase and antiphase active bending, we fix the swimmer in an oncoming uniform flow of speed $U$  and we compute the hydrodynamic drag forces in each case. In Figure~\ref{fig:compare2}, we report the drag force, input power, and circulation in the wake of the fixed swimmer.
Compared to rigid flapping, antiphase flexion reduces instantaneous drag, power, and circulation, while in-phase flexion increases all three quantities. Reduced drag implies lower thrust requirement during steady state swimming, which provides another perspective for understanding the improved performance of antiphase flexion.  

We next examine the period-average performance of actively bending swimmers as a function of anterior-to-posterior phase difference $\phi$. In Figure~\ref{fig:mix}(A), we consider the case of free swimming, we fix the flexion ratio $l/L=0.7$ and flexion amplitude $A=20^\circ$, and we plot the swimming speed $U$, input power $P$ and efficiency $\eta$ versus $\phi$, all scaled by the corresponding values of a rigidly flapping swimmer $U_{\rm rigid}$, $P_{\rm rigid}$, $\eta_{\rm rigid}$, respectively.
We find that active bending is always beneficial in terms of enhanced speed relative to rigid flapping, albeit at an increased power requirement. Importantly, as the anterior-to-posterior bending changes from in-phase flexion to flexion at a phase lag, the scaled speed increases and the scaled power requirement decreases. Optimal performance occurs at $\phi=0.9$ and $\phi=0.8$ in terms of maximum swimming speed and minimum input power and maximum efficiency, respectively. In Figure~\ref{fig:mix}(B), we fix the swimmer in oncoming flow of uniform speed $U$ and compute the scaled drag force $D$, input power $P$ and efficiency $\eta$ as a function of $\phi$ scaled by the corresponding values of a fixed rigid flapper. 
Analogous to the free swimmer, drag and input power are minimal at $\phi=0.8$. Taken together, the results imply that antiphase active flexion is near optimal for enhancing speed and efficiency and reducing drag force and power requirement.

To complete this analysis, we also explored the effect of the flapping parameter $M$ on the swimming performance. We found that for a fixed phase, the swimming speed and efficiency change monotonically with $M$ with maximum speed and minimum efficiency as $M\to 1$; see Figure~\ref{fig:appendix_M} in the Appendix. That is, reversing the bending direction with a quick flicker improves speed at the cost of decreasing efficiency.

%----
\begin{figure*}
\centering
\includegraphics[scale=1]{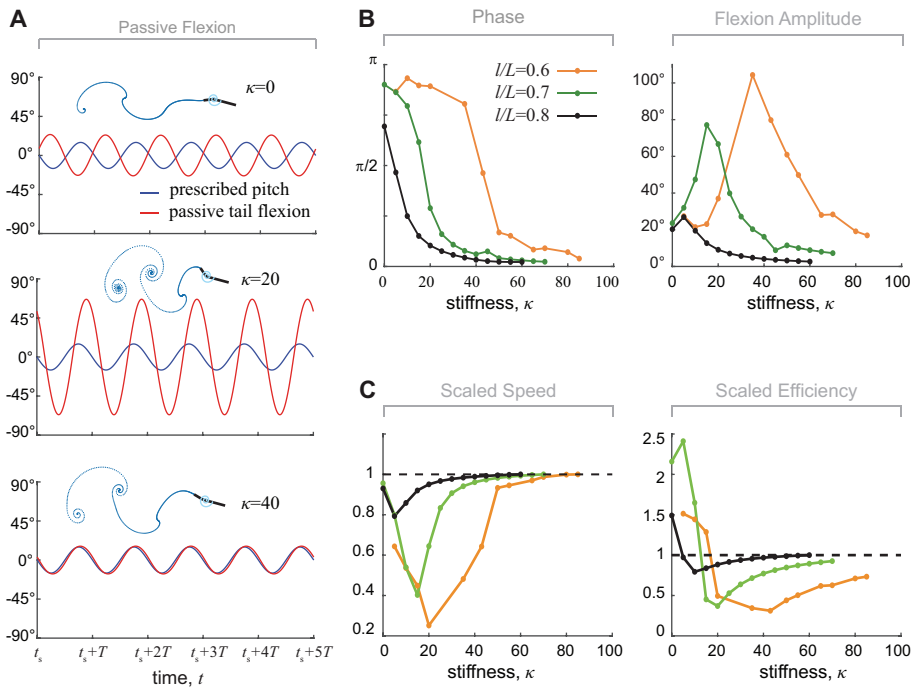}
\caption[]{\footnotesize \textbf{Passive anterior-to-posterior bending of free swimmers.} (A) Top to bottom: passive flexion angle $\alpha$ of posterior end at zero stiffness ($\kappa=0$),  moderate stiffness ($\kappa=20$), large stiffness ($\kappa=40$). The anterior link is pitching at $a=15^\circ$. The damping ratio is set to $c=1$, and flexion ratio to $l/L=0.7$. (B) The bending parameters (phase $ \phi$ and maximum flexion angle $A$). (C) Swimming speed $U$ and propulsion efficiency $\eta$ as a function of stiffness $\kappa$ for three flexion ratios $l/L=0.6$, $0.7$, $0.8$ reported in orange, green and black, respectively. The results are shown after the swimmers have reached steady state $t_{\rm{s}}=25T$. The dissipation time is set to be $T_{\rm{diss}}=\sqrt{2.09}T$. 
}
\label{fig:passive}
\end{figure*}
%----

Is active bending necessary for obtaining this enhancement in swimming speed and efficiency over rigid flapping? To address this question, we examine the free swimming of a passively bending swimmer, where the posterior end flaps passively under the effect of hydrodynamic forces and moments generated by the pitching motion of the anterior section. Elastic forces due to a spring of stiffness $\kappa$ located at the flexion point are also at play. In Figure~\ref{fig:passive}(A), we keep all parameter values the same as those used for the actively bending swimmer, and, from top to bottom, we report the flapping motions of the anterior and posterior ends for stiffness values $\kappa=0$, $20$, and $40$. At zero stiffness, flexion introduces no restoring forces and moments. The posterior part rotates antiphase relative to the flapping motion of the anterior part, at an amplitude comparable to the anterior pitching amplitude. The associated wake, shown in Figure~\ref{fig:flowfield}(D) and represented by the free vortex sheet in the inset of Figure~\ref{fig:passive}(A), shares similar features to the wake obtained during antiphase active flexion. At moderate spring stiffness $\kappa=20$, the flexion amplitude increases ($\alpha_{\max} \approx 80^\circ$), and the wake also exhibits larger lateral dispersion. 
At large stiffness $\kappa=40$, the  posterior part rotates in-phase with the anterior part at the same flapping amplitude in a way reminiscent to rigid flapping; as reported in \cite{Dai2012} for flexibble pitching plates. 

The relative motion of the posterior part is close to a sinusoidal function for all stiffness values $\kappa$. Therefore, for each $\kappa$ value, we fit $\alpha(t)$  by a sine function $\alpha(t)=A \sin (\omega t- \phi)$ using a 
% trust region reflective 
standard algorithm~\citep{More1983}. For all fitting, we have at least $95\%$ confidence and frequency $\omega \approx 2\pi$, thus ensuring  convergence of the fitting. In Figure~\ref{fig:passive}(B), we report the fitted flexion amplitude $A$ and phase difference $\phi$ as a function of spring stiffness $\kappa$ for three distinct flexion ratios $l/L =0.6$, $0.7$ and $0.8$. We find that for all $l/L$, as stiffness $\kappa$ increases, the phase difference ${\phi}$ decreases monotonously from $\pi$ to $0$ implying that the posterior flapping motion changes from antiphase to in-phase. The maximum flexion angle ${A}$ first increases with increasing $\kappa$, then decreases to nearly zero at large stiffness implying rigid flapping of both anterior and posterior ends. 

%----
\begin{figure*}
\centering
\includegraphics[scale=1]{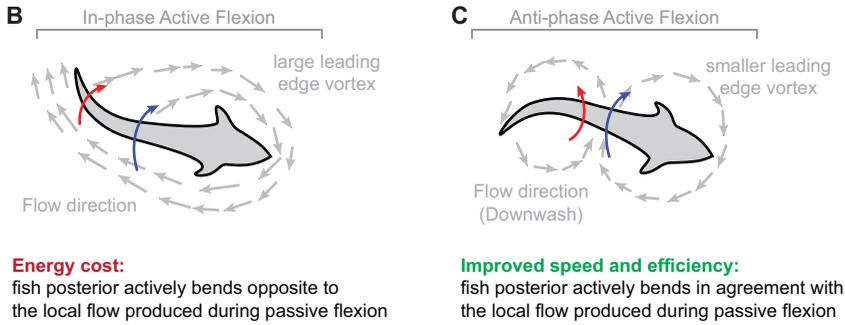}
\caption[]{\footnotesize \textbf{Active bending in agreement with passive hydrodynamics  improves swimming performance.} 
 Illustration of the interaction with the flow field for (A) in-phase and (B) anti-phase active flexion. Maximum improvement in swimming performance occurs when the tail benefits from flows created by the anterior portion of the fish body. 
Light grey arrows represent the flow direction. Red arrows and blue arrows represent the flapping direction of the posterior and anterior sections of the fish body, respectively. 
}
\label{fig:cartoon}
\end{figure*}
%---

We compute the associated swimming speed and efficiency for each stiffness value $\kappa$ and we scale the results by those of a rigid swimmer; see Figure~\ref{fig:passive}(C). Clearly, the swimmer with passive flexion never surpasses the swimming speed of a rigid swimmer. At very low stiffness ($\kappa \approx 0$), passive flexion results in a swimming speed close to that of rigid flapping while doubling the swimming efficiency. The increase in swimming efficiency at small stiffness comes purely from a decrease in power requirement compared to rigid flapping. This is in contrast to active flexion where the enhancement in swimming speed and efficiency noted in Figure~\ref{fig:mix} comes at an increase in power requirement relative to rigid flapping. 
As $\kappa$ increases, the scaled swimming speed  and propulsion efficiency decrease, indicating that moderate flexibility is detrimental to both speed and efficiency. For large $\kappa$, the speed and efficiency converge to the same speed and efficiency as the rigid swimmer, consistent with the results of~\cite{Dai2012}. We repeat this analysis for three flexion ratios $l/L=0.6$, $0.7$, $0.8$. The scaled speed seems to increase monotonically with increasing $l/L$, but the scaled efficiency seems to peak at $l/L = 0.7$ but only for a range of small $\kappa$ values. 
These findings imply that, unlike flapping insect wings~\citep{Ellington1985,Huang2015a}, restoring elastic forces seem to be detrimental to swimming performance. Swimming efficiency peaks at low stiffness values when the restoring spring forces are weak and the posterior end is driven passively by the fluid forces. 
%----
\begin{figure*}
\centering
\includegraphics[scale=1]{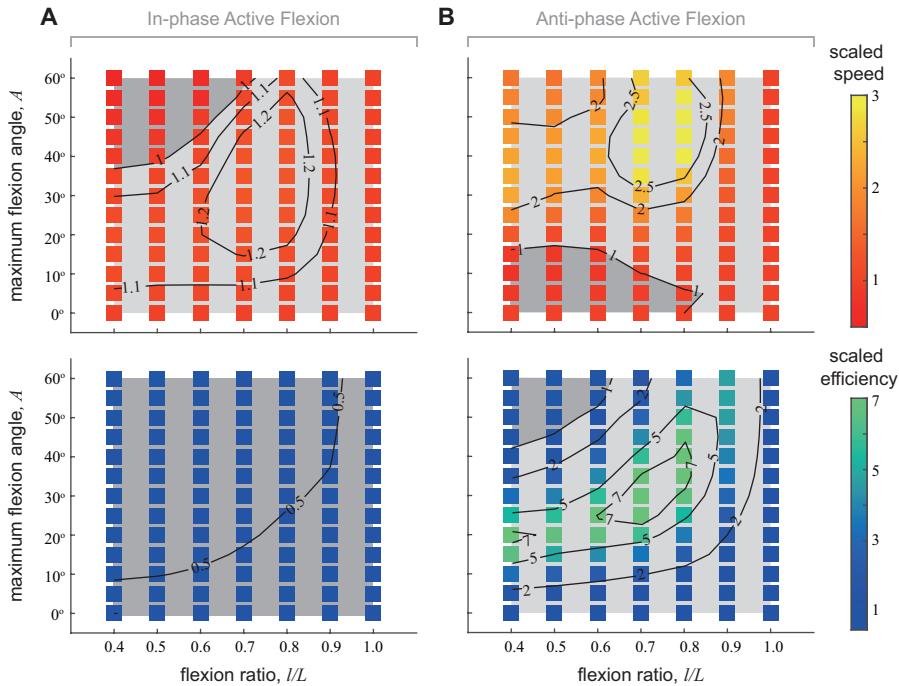}
\caption[]{\footnotesize \textbf{Swimming performance of actively flexion swimmer scaled by performance of rigid swimmer.} Average speed and efficiency versus flexion ratio $l/L$ and flexion angle $A$ for (A) in-phase flexion ($\phi=0$) and (B) antiphase flexion ($\phi=\pi$). The amplitude for the proximal part and the elliptic modulus are $a=15^\circ$ and $M=0.9$. Dark grey areas indicate regions of diminished performance while light gray areas indicate improved performance over a rigid swimmer. The dissipation time is set to be $T_{\rm{diss}}=1.65T$. 
}
\label{fig:colormap}
\end{figure*}
%----

Could the swimmer learn from passive flexion to improve its performance by bending actively in a way that exploits the hydrodynamic forces generated naturally during passive flexion? The results in Figures~\ref{fig:flowfield}-\ref{fig:mix} for actively bending swimmers suggest that maximum benefit occurs for near anti-phase flexion, whereas the results in Figure~\ref{fig:passive} show that maximum efficiency for passively bending swimmers occurs for zero stiffness $(\kappa = 0)$ for which the posterior bends anti-phase. Importantly, the main features of the instantaneous flow during antiphase active flexion (Figure~\ref{fig:flowfield}C) are also observed in passive bending at zero stiffness (Figure~\ref{fig:flowfield}D). We thus posit that active bending is most beneficial when the swimmer actively beats its tail in a direction that takes advantage of the natural flows that arise during passive bending.
To test this hypothesis, we define a flexion agreement parameter $Z$ that aims to relate passive and active bending. Starting from a swimmer bending passively at zero spring stiffness $\kappa=0$ (Figures~\ref{fig:flowfield}D and~\ref{fig:passive}A), we assume a hypothetical posterior that is actively flapping about the flexion point of the swimmer at a relative angle  $\alpha = A \sin(2\pi t - \phi)$. We compute the fluid velocity $\mathbf{u}(s,t)$ induced by the passively-bending swimmer along the hypothetical posterior section that is bending actively (here, $s$ is the distance from the flexion point). The flexion agreement parameter $Z$ is given by
\begin{equation}
Z=\frac 1 T \frac 1 {L-l}\int_0^T{\int_0^{L-l}{\mathbf{u}(s,t) \cdot \mathbf{v}(s,t)ds}dt},
\label{eq:flexion_agreement}
\end{equation}
where $\mathbf{v}(s,t) = s \dot{\alpha} \,\mathbf{n}$ is the relative velocity of the hypothetical actively-flapping posterior. Positive values of the flexion agreement parameter imply a beneficial interaction between  the flow generated during passive flexion and the velocity of the hypothesized posterior during active flexion, whereas negative values indicate a detrimental one.

In Figure~\ref{fig:mix}(A), we set $A$ to be equal to the maximum flexion angle of the passively-bending swimmer, and we vary $\phi$ from $0$ to $\pi$. We find that the agreement parameter $Z$, normalized by its maximum value, is largest for antiphase active flexion and smallest near in-phase active flexion.  This result indicates that antiphase active flexion matches the local flow created during passive flexion best, whereas in-phase active flexion acts opposite to these flows. That is, the swimmer during antiphase flexion can utilize better the flow field generated by the pitching motion of its anterior section, and thus it can achieve higher swimming speed and efficiency compared to in-phase flexion.
%, as illustrated schematically in Figure~\ref{fig:cartoon}(B,C). 

To emphasize the effect of the interaction between the flow field and kinematics of active flexion on swimming performance,
we schematically summarize the two cases of in-phase and antiphase flexion in Figure~\ref{fig:cartoon}. %reported in 
%Figure~\ref{fig:colormap} and used to produce the regions of enhanced perfomance in 
%Figure~\ref{fig:cartoon}(A).
During in-phase active flexion, the flow field is characterized by a strong leading edge vortex around the anterior section of the fish and large lateral forces. Further, the posterior part bends opposite to the local flow field of a passively bending swimmer. 
During antiphase active flexion, the leading edge vortex is smaller and it is followed by a counter-rotating vortex around the fish mid-section such that the posterior part is moving in synchrony with the downwash flow induced by this counter-rotating vortex. The flow field is helping the motion of the posterior end. These results indicate that active body deformation that are in agreement with local flows produced during passive deformations are more advantageous for enhancing swimming speeds and efficiencies. 
%They also emphasize the effect of the interaction between the flow field and kinematics of active flexion on the swimming performance reported in 
%Figure~\ref{fig:colormap} and used to produce the regions of enhanced perfomance in 
% Figure~\ref{fig:cartoon}(A).

%----
\begin{figure*}
\centering
\includegraphics[scale=1]{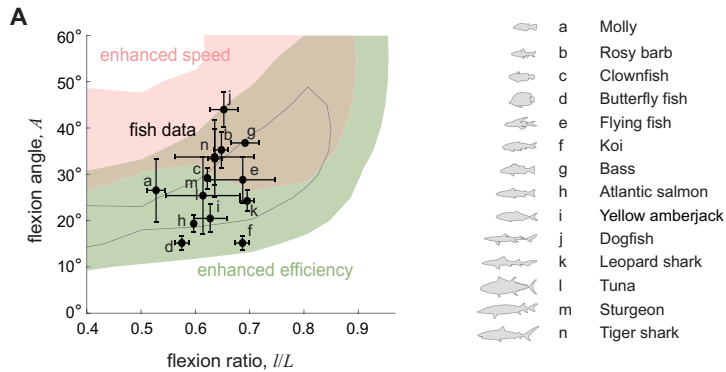}
\caption[]{\footnotesize \textbf{Relation to fish swimming behavior. } 
% Active bending in agreement with passive bending enhances swimming speed and efficiency.
%\textcolor{blue}{this sentence need to be changed.}
(A) Comparison of fish flexion parameters (black dots, cite from \cite{Lucas2014}) and regions of optimal performance predicted by our model (anti-phase active flexion swimmer): pink region corresponds to 200\% enhancement in swimming speed, and green region corresponds to 300 \% enhancement in swimming efficiency, both compared to a swimmer of the same total length rigidly flapping with no flexion. Overlap of the two regions is indicated in beige. The contour grey line encloses a region of 600\% enhancement in efficiency. 
%(B-C) Illustration of the interaction with the flow field for in-phase and anti-phase flexion. Maximum energy savings occur when the tail benefits from flows created by the leading portion of the fish body. 
%Light grey arrows represent the flow direction. Red arrows and blue arrows represent the flapping direction of the posterior and anterior sections of the fish body, respectively. 
}
\label{fig:compareFish}
\end{figure*}
%---

Lastly, we explore the effect of maximum flexion angle $A$ and flexion ratio $l/L$ on the period-average values of the swimming speed and efficiency for both in-phase and antiphase active flexion. Specifically, we examine the range $l/L \in [0.4, 1.0]$ and $A \in [0^\circ, 60^\circ]$ for $\phi = 0$ and $\phi = \pi$.  In Figure~\ref{fig:colormap}, we report the period-average values normalized by the corresponding values for a rigid swimmer with pitching amplitude equal to the anterior part amplitude.
We highlight in light and dark grey respectively the regions in the parameter space where the flexible swimmer outperforms and underperforms the rigid swimmer. The swimmer with in-phase flexion swims slower than the rigid swimmer for small flexion ratios ($l/L < 0.7$) and high flexion amplitudes ($A >40^\circ$), and swims faster than rigid swimmer otherwise. This swimmer, however, is always less efficient than the rigid swimmer for reasons explained previously. In Figure~\ref{fig:colormap}(B), for most parameter values, the antiphase swimmer outperforms the rigid swimmer in terms of swimming speed and efficiency. Note that the region with the highest swimming speed advantage lies in $l/L \in [0.6, 0.8]$ and $ A \in [30^\circ,  60^\circ]$, and the region with the highest efficiency advantage lies in $l/L \in [0.6, 0.8]$ and $ A \in [25^\circ, 50^\circ]$.

We compare the regions of highest swimming speed and efficiency obtained during antiphase flexion in Figure~\ref{fig:colormap}(B) to the design parameters of biological fish reported in Figure~\ref{fig:biodata}. In Figure~\ref{fig:compareFish}, we plot the flexion angle $A$ as a function of the flexion ratio $l/L$ for the fish data in Figure~\ref{fig:biodata} and we superimpose on this design space the regions of 200\% enhancement in speed and 300\% enhancement in efficiency from Figure~\ref{fig:colormap}(B).
As shown in Figure~\ref{fig:compareFish}, there is significant overlap between these regions of improved performance and the biological data. Indeed, all biological data lie within the region of improved efficiency. % and some in the region of faster swimming speed. 

Many of these fish are known to exhibit migratory behavior that requires efficient swimming. Even baby clownfish are reported to migrate over long distances~\citep{Simpson2014}. Tuna can cover $7600$ km  in one traveling phase~\citep{Itoh2003} and tiger shark are capable of traveling long distances in short time~\citep{Simpfendorfer2009}.  Our results in Figure~\ref{fig:compareFish} are consistent with these facts: tiger shark and clownfish lie in the intersection region of improved speed and efficiency, and tuna lies in the region with highest increase in efficiency. On the hand, butterflyfish, who are only known to migrate over short distances during spawning \citep{Yabuta1997}, and Koi, for which there is no evidence of migration, lie in the regions characterized by smaller increase in speed and efficiency.

\section{Conclusion}\label{sec:con}

We analyzed the swimming performance of flapping swimmers undergoing active and passive deformations. Whereas fish exhibit a variety of swimming modes~\cite{Sfakiotakis1999}, we simplified body deformations to account for only anterior-to-posterior bending, with one degree of freedom describing the relative rotation between the two sections. We explored the effects of morphological and kinematic parameters on the swimming speed and efficiency. 

We found that passive body bending, at negligible body stiffness and minor elastic forces, caused anterior-to-posterior antiphase flexion. This antiphase flexion is dictated by the flow physics and causes the swimmer's morphology to get more streamlined compared to rigid flapping with no flexion, thus creating leaner wakes that reduce drag and power requirement and increase efficiency. While drag reduction is desirable for improved efficiency, passive bending also reduced thrust production, thus diminishing swimming speed.  Interestingly, restoring elastic forces at moderate stiffness values seemed detrimental to both swimming speed and efficiency. These findings are consistent with the hypothesis that for maximum efficiency, the fish tail and posterior body should flex like water, exhibiting little or no resistance to the flows generated by the flapping motion of the anterior portion of the body. This hypothesis could explain how anesthetized fish, with no muscle activity, placed in periodic wakes generate oscillatory body deformations that allow the fish to swim upstream~\citep{Beal2006}.

We also found that a swimmer that actively creates antiphase anterior-to-posterior bending enjoyed the same benefits of leaner wake and reduced drag as a passively bending swimmer while mitigating the reduction in thrust and swimming speed.
To quantify the hydrodynamic mechanisms leading to this improved swimming performance, we introduced a flexion agreement parameter that compares the active flexion velocity of the swimmer's posterior to the local flow velocity during passive flexion. We found that during active in-phase flexion, the posterior beats opposite to the local flow that would naturally arise during passive flexion, leading to a negative flexion agreement parameter, and thus lower swimming speed and efficiency. 
During active antiphase flexion, the posterior flaps consistently with the local flow, leading to a positive flexion agreement parameter, and improved speed and efficiency. 

These findings suggest tremendous versatility in swimming performance, even when accounting only for coarse anterior-to-posterior bending motions. They indicate that fish can readily and fluidly transition from efficient (passive bending) to fast (active bending) by actively beating their tail in agreement with the local flow generated during passive bending. 

To explore the role of flow physics on the convergent bending rules of \citep{Lucas2014}, we examined the effect of flexion ratio and maximum flexion angle on the performance of swimmers undergoing active antiphase anterior-to-posterior bending.  We found an optimal region in this design space that simultaneously enhances swimming speed and efficiency. Importantly, we found that this region has a significant overlap with the fish bending parameters reported in \citep{Lucas2014}; see Figure~\ref{fig:cartoon}(A).  In fact, all biological data fell in the region of enhanced efficiency predicted by our model.

Taken together, our results have two major implications on understanding the role of body bending in fish swimming. They are consistent with the hypothesis that fish that actively bend their bodies in a fashion that exploits the local hydrodynamics can at once improve speed and efficiency. They also support the hypothesis that flow physics could have provided a selective force for driving the evolution of fish bending patterns.

Beyond the trade-offs between speed and efficiency explored here, our model could be generalized in future studies to examine transient swimming maneuvers such as turning. \cite{Pollard2019} proposed that a passive posterior not only improves efficiency, but also improves fish maneuverability compared to a rigid body. \cite{Drucker2000} pointed out that although large lateral forces are detrimental to fish swimming speed, they can improve fish maneuverability. Our model predicts large lateral forces during in-phase active flexion, suggesting that this bending pattern, while not optimal for forward swimming, could be beneficial for turning motions. These considerations, as well as models of higher fidelity to the fish biomechanics and fluid environment, will be explored in future research. 

Finally, our finding that active body bending and tailbeat patterns that match local flow velocities that would be produced naturally by anterior sections of the fish body could lead to improved performance and energy savings might have important implications on understanding the mechanisms driving body and caudal fin deformations in swimming~\citep{Ramakrishnan2011,Bozkurttas2007}, schooling~\citep{Heydari2020,Becker2015,Li2020}, and navigating ambient unsteady flows~\citep{Liao2003}.

% \paragraph{Acknowledgment.} The work is partially supported by the National Science Foundation (NSF) through the grant NSF CBET 21-ZZZ. EK also acknowledges support from the Office of Naval Research through the grants XXX and YYY.

\bibliography{jfm}
\bibliographystyle{jfm}

\appendix
\renewcommand\thefigure{\Alph{section}}
\section{Vortex sheet model}
\label{app:VSM}
The coupled fluid-structure interaction between the two-link swimmer and the surrounding fluid is simulated using an inviscid vortex sheet model. 
In viscous fluids, boundary layer vorticity is formed along the sides of the swimmer, and it is swept away at the swimmer's tail to form a shear layer that rolls up into vortices. 
In the vortex sheet model, the swimmer is approximated by a bound vortex sheet, denoted by $l_\b$, whose strength ensures that no fluid flows through the rigid plate, and the separated shear layer is approximated by a free regularized vortex sheet $l_\w$ at the trailing edge of the swimmer.
The total shed circulation $\Gamma$  in the vortex sheet is determined so as to satisfy the Kutta condition at the trailing edge, which is given 
in terms of the tangential velocity components above and below the 
bound sheet and ensures that the pressure jump across the sheet vanishes at the trailing edge. 

To express these concepts mathematically, it is convenient to use the complex notation $z = {\rm x} + \ii {\rm y}$, where $\ii = \sqrt{-1}$ and $({\rm x},{\rm y})$ denote the components
of an arbitrary point in the plane. 
The bound vortex sheet $l_\b$ is described by its position $z_\b(s,t)$ and strength $\gamma(s,t)$, 
where $s\in[0,L]$ denotes the arc length along the sheet $l_\b$.
The separated sheet $l_\w$ is described by its position $z_\w(\Gamma,t)$, $\Gamma\in[0,\Gamma_\w]$ where $\Gamma$ is the Lagrangian circulation around the portion of the separated sheet between its free end in the spiral center and the point $z_\w(\Gamma,t)$. 
The parameter $\Gamma$ defines the vortex sheet strength $\gamma=d\Gamma/ds$.

By linearity of the problem, the complex velocity $w(z,t) = u(z,t) - \mathrm{i}v(z,t)$ 
is a superposition of the contributions due to the  bound and free vortex sheets 
%----
\begin{equation}
w(z,t)= w_\b(z,t) + w_\w(z,t).
\label{flowfield}
\end{equation}
%-----
In practice, the free sheet $l_\w$ is regularized
using the vortex blob method to prevent the growth of the Kelvin-Helmholtz 
instability. The bound sheet $l_\b$ is not regularized in order to preserve the invertibility of the map between the sheet strength and the normal velocity along the sheet. The velocity components $w_\b(z,t)$ and $w_\w(z,t)$  induced by the bound and free vortex sheets, respectively, are given by
%---
\begin{equation}
\begin{aligned}
w_\b(z,t) = 
\int_{0}^{L} K_o(z-z_{b}(s,t))\gamma(s,t)\, ds, \quad
w_\w(z,t) =  
\int_0^{\Gamma_\w}K_{\delta}(z-z_\w(\Gamma,t))\, d\Gamma,
\end{aligned}
\label{fluidvelo}
\end{equation}
where $K_\delta$ is the vortex blob kernel, with regularization parameter $\delta$,
\begin{equation}
K_\delta (z) = \frac{1}{2\pi \mathrm{i}}\frac{\overline{z}}{|z|^2+\delta^2}, \qquad \overline{z} = x - \ii y.
\end{equation}
If $z$ is a point on the bound sheet for which $\delta = 0$, $w_\b$ is to be computed in the principal value sense.  
%----------
\begin{figure*}
\label{fig:appendix}
\centering
\includegraphics[scale=1]{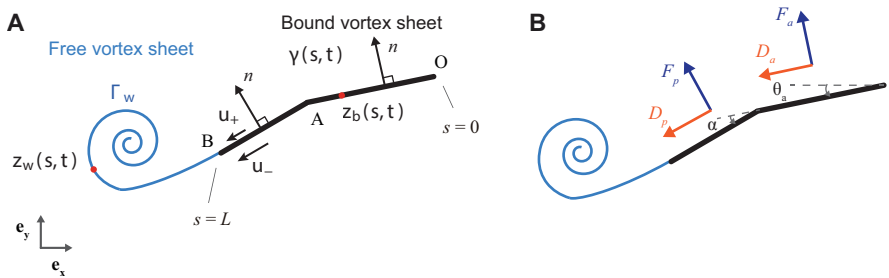}
\caption[]{\footnotesize (A) Schematic of the vortex sheet model for a two-dimensional bending swimmer. (B) Depiction of the different hydrodynamic forces acting on the swimmer. }
\end{figure*}
%----------
The position of the bound vortex sheet $z_\b$
is determined from the plate's flapping ($\theta_a(t), \theta_p(t)$) and swimming $x(t)$ motions.
The corresponding sheet strength $\gamma(s,t)$
is determined by imposing the no penetration boundary condition 
on the plate, together with conservation of total circulation. 
The no penetration boundary condition is given by
%--------
\begin{equation}
\Real \left[ w n\right]_{z_b}= 
\Real \left[  w_{\rm swimmer} n\right],
\end{equation}
%-----
where 
\begin{equation}
 n=\left\{\begin{array}{lr}
-\sin \theta_a+ \ii \cos  \theta_a, & s \in [0,l],  \\
-\sin \theta_p+ \ii \cos \theta_p, & s \in [ l,L],
\end{array}\right.
\end{equation}
%--------
and 
%--------
\begin{equation} w_{\mathrm{swimmer}}=\left\{\begin{array}{lr}
\dot{x} - \ii \dot{y}-\mathrm{i}\dot{\theta_a}\left[\bar{z}_b - ({x}- \ii {y})\right],   s \in [0,l] , \\
\dot{x} - \ii \dot{y}-\mathrm{i}\dot{\theta_a}\left[\bar{z}_b - ({x}- \ii {y})\right]-\mathrm{i}\dot{\alpha}\left[\bar{z}_b - ({x}_{A}- \ii {y}_{A})\right], & s \in [ l,L].
\end{array}\right. 
\end{equation}
%--------
Conservation of the fluid circulation implies that
$\int_{l_b} \gamma(s,t) ds + \Gamma_\w (t) = 0$.

The circulation parameter $\Gamma$ along the free vortex sheet $z_\w(\Gamma,t)$
is determined by the circulation shedding rates $\dot\Gamma_\w$, according to 
the Kutta condition, which states that the fluid velocity at the trailing edge is finite and tangent to the flyer.
The Kutta condition can be obtained from the Euler equations by enforcing that, at the trailing edge, the difference in pressure 
across the swimmer is zero. To this end, we integrate the balance of momentum equation for inviscid planar flow
along a closed contour containing the vortex sheet and trailing edge,
%---
\begin{equation}
[p]_\mp(s) = p_-(s) - p_+(s) =  - \frac{d\Gamma(s,t)}{dt} - \frac{1}{2}(u_-^2 - u_+^2),
\label{pressure}
\end{equation}
%---
where $\Gamma(s,t) = \Gamma_\w+\int_{0}^{s} \gamma(s',t)ds'$, $0 \le s \le L$, 
is the circulation within the contour and $p_{\mp}(s,t)$ 
and $u_\mp(s,t)$  denote the limiting pressure and tangential 
slip velocities on both sides of the swimmer. 
Since the pressure difference across the free sheet is zero, 
it also vanishes at the trailing edge by continuity, which implies that
%----
\begin{equation}
\dot{\Gamma}_{\w}=-\frac{1}{2}(u_-^2 - u_+^2)|_{s=L}.
\label{sheddingrates}
\end{equation}
The values of
$u_-$ and $u_+$ 
are obtained from the average tangential velocity component and 
from the velocity jump at the trailing edge, given by the sheet strength, evaluated at $s=L$
%---
\begin{equation}
\overline{u}={\frac{u_+ + u_-}2}=\Imag[(w-w_{\rm swimmer})n]~,
\qquad 
u_--u_+= \gamma .
\label{veloedge}
\end{equation}
%---
Once shed, the vorticity in the free sheet moves with the flow. 
Thus the 
parameter $\Gamma$ assigned to each particle $z_\w(\Gamma,t)$
is the value of $\Gamma_\w$ at the instant it is shed from the trailing edge.
The evolution of the free vortex sheet $z_\w$ is obtained by 
advecting it in time with the fluid velocity,
%---
\begin{equation}
\dot{\bar{z}}_{\w} = w_\w(z_{\w},t) +w_\b(z_{\w},t).
\label{freesheets}
\end{equation}
%---

\section{Forces and moments}
\label{app:forces}

The hydrodynamic forces $F_a$ and $F_p$  acting on the anterior and posterior parts of the swimmer, respectively, are given by 
%---
\begin{equation}
\begin{split}
%\begin{array}{l}
F_a= F_{ax} + \ii F_{ay}=\int_{0}^l n[p]_\mp d s, \qquad 
F_p= F_{px} + \ii F_{py}=\int_{l}^L n[p]_\mp d s ,
%\end{array}
\end{split}
\label{force}
\end{equation}
%---
The hydrodynamic moment $M_a$ acting on anterior part of the swimmer about its leading edge and  the hydrodynamic moment $M_p$ acting on the posterior part of the swimmer about the flexion point are given by
%---
\begin{equation}
\begin{split}
%\begin{array}{l}
M_a = \int_{0}^l{[p]_\mp s ds}, \qquad 
M_p =\int_{l}^L{[p]_\mp (s-l)ds}.
%\end{array}
\end{split}
\label{eq:moment}
\end{equation}
%--
Note that the components  $F_{ax}, F_{ay}$ and $ F_{px}, F_{py}$  can be written explicitly as 
%---
\begin{equation}
\label{eq:force}
\begin{split}
\begin{array}{l}
 F_{ax}=\int_{0}^l{[p]_\mp (-\sin \theta_a)ds}, \qquad  F_{ay}=\int_{0}^l{[p]_\mp \cos\theta_a ds}, \\
 F_{px}=\int_{l}^{L}{[p]_\mp (-\sin \theta_p)ds}, \qquad  F_{py}=\int_{l}^{L}{[p]_\mp \cos \theta_p ds}.  
\end{array}
\end{split}
\end{equation}
%---
where $\theta_p = \theta_a + \alpha$ and $\alpha$ is the flexion angle.

The total hydrodynamic force acting on the swimmer due to the pressure difference across the swimmer is given by
%---
\begin{equation}
\begin{split}
\begin{array}{l}
F=F_x + \ii F_y %= \int_{l_b} n[p]_\mp d s , 
\end{array}
\end{split}
\label{eq:force_total}
\end{equation}
%---
where the components $F_x$ and $F_y$ are 
%---
\begin{equation}
\begin{split}
\begin{array}{l}
F_x = F_{ax} + F_{px} = \int_{0}^{l}{[p]_\mp (-\sin \theta_a)ds}+\int_{l}^{L}{[p]_\mp (-\sin \theta_p)ds}, \\
F_y = F_{ay} + F_{py} =  \int_{0}^{l}{[p]_\mp \cos \theta_a ds}+\int_{l}^{L}{[p]_\mp \cos \theta_p ds},  
%\\
%M = \int_{0}^l{[p]_\mp s ds}+\int_{l}^{L}{[p]_\mp (s-l+l \cos \alpha)ds},
\end{array}
\end{split}
\label{eq:force_total_components}
\end{equation}
%---
%where, $\alpha=\theta_p-\theta_a$ is the flexion angle. 
The total hydrodynamic moment acting on the swimmer about its leading edge is given by
%---
\begin{equation}
\begin{split}
\begin{array}{l}
M = \int_{0}^l{[p]_\mp s ds}+\int_{l}^{L}{[p]_\mp (s-l+l \cos \alpha)ds}.
\end{array}
\end{split}
\label{eq:moment_total}
\end{equation}
%---

We introduce a drag force $D$ that emulates the effect of skin friction due to fluid viscosity. This force is based on the Blasius laminar boundary layer theory as implemented by \cite{Fang2016} in the context of the vortex sheet model. Blasius theory provides an empirical formula for skin friction on one side of a horizontal plate of length $L$ placed in fluid of density $\rho_f$ and uniform velocity $U$.
In dimensional form, Blasius formula is $D = - \frac{1}{2} \rho_\f L (c_\f)U^2$, where the skin friction coefficient $C_\f = {0.664}/{\sqrt{\rm Re}}$ is given in terms of the Reynolds number $\textrm{Re} = {\rho_f U L}/{\mu}$. 
Substituting back in the empirical formula leads to $D = - C_d U^{3/2}$, where $C_d = 0.664 \sqrt{ \rho_f \mu (L)}$.
Following \cite{Fang2016}, we write a modified expression of the drag force for a swimming plate
%--
\begin{equation}
D =  C_d (\overline{U}_+ ^{3/2} + \overline{U}_- ^{3/2}),
\end{equation}
%--
where $\overline{U}_{\pm}$ are the spatially-averaged tangential fluid velocities on the upper and lower side of the plate, respectively, relative to the swimming velocity $U$,
%--
\begin{equation}
\overline{U}_{\pm}(t) = \frac{1}{L} \int_{0} ^{L} u_{\pm} (s,t)ds - U.
\end{equation}
%--
We estimate $C_d$ to be approximately $0.04$ in the experiments of \cite{Ramananarivo2016}.

The equation of motion governing the free swimming $x(t)$ is given by Newton's second law
%--
\begin{equation}
m \ddot{x}=  F_x - D_x, 
\label{eq:eom}
\end{equation}
%--
where $D_x$ is  the $x$-component of the drag force $D$. When the swimmer bends passively, the relative rotation $\alpha(t) = \theta_p -\theta_a$ of the posterior end is not prescribed a priori and follows from the physics of fluid-structure interactions. Considering that the rotational joint at the flexion point is equipped with a torsional spring of stiffness $\kappa$ and damping coefficient $c$, we write the equation governing the relative rotation of the posterior link
%--
\begin{equation}
I_{p}(\ddot{\theta}_a + \ddot \alpha) +  c \dot{\alpha} + \kappa \alpha =  M_{p} + M_{\rm inertia}, 
\label{eq:eomalpha}
\end{equation}
%--
%\ek{double check the above equation and correct the equation that is in the main text (in section 2) accordingly}
where $I_{p}$ is the moment of inertia of the posterior link about the flexion point, and $M_{\rm{inertia}}$ is an inertial moment acting on the posterior link due to the free motion of the flexion point. Namely,
%----
\begin{equation}
M_{\rm intertia} = m_p \textrm{Im} \left[ -  z_{\rm A} \bar{a}_A \right],
\end{equation}
%----
%\ek{double check that the expression in the square bracket is a cross product between two vectors $z_A$ and $a_A$ expressed in complex notation: checked, it is correct} 
where $m_p=\rho e (L-l)$ is the mass of the posterior link, $z_{\rm A} = \dfrac{L-l}{2}\left( \cos\theta_p + \ii \sin\theta_p\right)$ is the position of the flexion point relative to the mass center of the posterior link, and $\bar{a}_A$ is the complex conjugate of the acceleration ${a}_A$ at the flexion point. The latter is given by 
%----
\begin{equation}
a_A=(\ddot x -l \ddot \theta_a \sin \theta_a) -\ii l \dot \theta_a^2\sin \theta_a.
\end{equation}
%----

For a swimmer undergoing active flexion, the flapping motion and body bending are produced by two active moments $M_{\rm a}$ and $M_{\rm p}$ acting by the swimmer on the fluid about the leading edge $\rm O$ and the hinge $\rm A$, respectively. 
The power input by the swimmer to overcome the moment of all the hydrodynamic forces about the leading edge is given by
%---
\begin{equation}
\begin{split}
\begin{array}{l}
P(t)=\dot{\theta}_a M_a+\dot{\theta}_p (M_p+l |F_p| \cos \alpha ). 
\end{array}
\end{split}
\end{equation}
%---
For a swimmer with passive flexion, the input power is given by
%---
\begin{equation}
\label{eq:derive2}
\begin{split}
\begin{aligned}
P(t)=\dot{\theta}_a (M_a+l  |F_p| \cos \alpha-\kappa \alpha- c \dot \alpha)
\end{aligned}
\end{split}
\end{equation}
%---
Note that the skin drag does not contribute to input power. 

%----------
\begin{figure*}
\label{fig:appendix_M}
\centering
\includegraphics[scale=1.2]{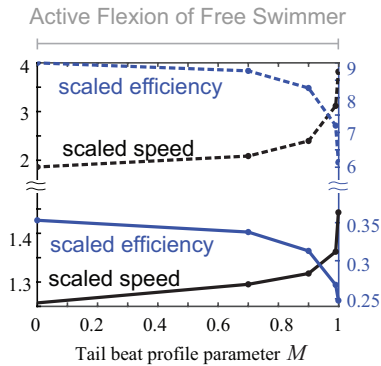}
\caption[]{\footnotesize \textbf{Swimming performance versus tail beat profile parameter} (elliptic modulus $M$). Swimming speed and efficiency of active flexion scaled by rigid flapping for $a=15^\circ$, $l/L=0.7$, $A=30^\circ$, respectively. As $M$ increases, the relative flapping of the posterior end goes from sinusoidal to square-like wave, with faster reversal of flapping direction during the flapping cycle. Dashed lines correspond to anti-phase active flexion and  and solid lined to in-phase flexion.}
\end{figure*}
%----------

\section{Numerical implementation}
\label{app:numerics}
The bound vortex sheet is discretized by $2n+1$ point vortices at $z_b(t)$ with strength $\Delta\Gamma=\gamma\Delta s$.
These vortices are located at Chebyshev points that cluster at the two ends of the swimmer.
Their strength is determined by enforcing no penetration at the midpoints between the vortices, together with conservation of circulation.
The free vortex sheet is discretized by regularized point vortices at $z_{\w}(t)$, that is released from the trailing edge at each timestep with circulation given by \eqref{sheddingrates}.
The free point vortices move with the discretized fluid velocity while the bound vortices move with the swimmer's velocity. 
%\textcolor{red}{
For the actively-bending swimmer, the discretization of equations (\ref{eq:eom}) and (\ref{sheddingrates}, \ref{freesheets}) yields a coupled system of ordinary differential evolution equations for the swimmer's position, the shed circulation, and the free vorticity, that is integrated in time using the 4th order Runge-Kutta scheme. For the passively-bending swimmer, the discretization of equation (\ref{eq:eomalpha}) is added to the coupled system of equations to simultaneously solve for the rotational motion of the posterior link relative to the anterior link.
% simultaneously with other physics quantities. } 
%\ek{This is not exactly complete! This is only true to active flapping. You need to discuss the equations you solve when flapping passively}
The details of the shedding algorithm are given in \citep{Nitsche1994}. 
The numerical values of the timestep $\Delta t$, the number of bound vortices $n$, and the regularization parameter $\delta$ are chosen so that the solution changes little under further refinement.

Finally, to emulate the effect of viscosity, we allow the shed vortex sheets to decay gradually by dissipating each incremental point vortex after a finite time $T_{\rm diss}$ ($T_{\rm diss}=1.65T$ for the swimmer with active flexion and $T_{\rm diss}=\sqrt{2.09}T$ for passive flexion) from the time it is shed into the fluid.
Larger $T_{\rm diss}$ implies that the vortices stay in the fluid for longer times, mimicking the effect of lower fluid viscosity. 
We refer the reader to \cite{Huang2018} for a detailed analysis of the effect of dissipation time on the hydrodynamic forces on a stationary and moving plate in the vortex sheet model. Details of the numerical validation in comparison to \cite{Jones2003} and \cite{Jones2005} are provided in \cite{Huang2016}.

\end{document}